\documentclass[prd,showpacs,11pt,nofootinbib,aps]{revtex4-1}
\usepackage{amsmath}
\usepackage{amssymb}
\usepackage[dvips]{graphicx}
\usepackage{hyperref}
\usepackage{url}
\usepackage[all]{xy}

\begin{document}

\title{Bouncing solutions in Rastall's theory with a barotropic fluid}
\author{G. F. Silva}\email{gabrielfsilva@gmail.com}\author{O. F. Piattella}\email{oliver.piattella@pq.cnpq.br}\author{J. C. Fabris}\email{fabris@pq.cnpq.br}\author{L. Casarini}\email{casarini.astro@gmail.com}\author{T. O. Barbosa}\email{taislaneoliveira@gmail.com}
\affiliation{Physics Department, Universidade Federal do Esp\'irito Santo, Vit\'oria, 29075-910, ES, Brazil}

\begin{abstract}
Rastall's theory is a modification of Einstein's theory of gravity where the covariant divergence of the stress-energy tensor is no more vanishing, but proportional to the gradient of the Ricci scalar. The motivation of this theory is to investigate a possible non-minimal coupling of the matter fields to geometry which, being proportional to the curvature scalar, may represent an effective description of quantum gravity effects. Non-conservation of the stress-energy tensor, via Bianchi identities, implies new field equations which have been recently used in a cosmological context, leading to some interesting results. In this paper we adopt Rastall's theory to reproduce some features of the effective Friedmann's equation emerging from loop quantum cosmology. We determine a class of bouncing cosmological solutions and comment about the possibility of employing these models as effective descriptions of the full quantum theory.
\end{abstract}

\maketitle

\section{Introduction}

The existence of an initial singularity in the standard cosmological model is one of the most important problems to be treated by a theory of gravity. This problem comes together with the existence of singularities in the spherical solutions of General Relativity (GR), connected with the notion of black holes. But, while in the black hole case the singularity is covered by an event horizon, being inaccessible to an external observer, the cosmological singularity does not have such kind of protection, and the emergence of the universe from a singular state renders the initial conditions of the universe quite arbitrary and, strictly speaking, unpredictable. The existence of singularities in General Relativity is related to the so-called energy conditions \cite{ellis} that any ordinary matter should satisfy.
\par
Classically, the initial cosmological singularity can be avoided by introducing ad-hoc fluids that violate the energy conditions. This may even imply to consider a phase where the total energy density is negative (violation of the weak energy condition) or a negative pressure such that $\rho + p < 0$ (violation of null energy condition). Another possibility is a modified theory of gravity, like scalar-tensor theories or non-linear Lagrangians. Quantum effects may lead to the appearance of negative energy states, as it happens in the Casimir effect \cite{moste}. Such fact gives the hope that the construction of a quantum theory of gravity may solve the singularities problem of General Relativity.
\par
One possible modification of the classical theory of gravity is given by Rastall's theory \cite{rastall1, rastall2}. In this theory the usual conservation law represented by the zero divergence of the energy-momentum tensor is
given up; instead, the divergence of the energy-momentum tensor is related to the Ricci curvature scalar. The main argument in favor of such proposal is that the usual conservation laws connected with the divergence of the energy-momentum tensor are tested only in the Minkowski space-time (more specifically, in a weak field limit). Even if Rastall's theory is incomplete, in the sense that it lacks a Lagrangian formulation (at least in the context of the Riemaniann geometry), it has an interesting feature: it reproduces in a quite phenomenological way, but in an explicit covariant form, a distinguishing property of quantum effects in gravity systems, the violation of the classical conservation laws \cite{birrell}. In this way, Rastall's theory may be viewed as a classical, phenomenological way to implement some concepts that are typical of a quantum theory of gravity. While Rastall's theory has already been studied in the context of the present evolution of the universe \cite{Fabris:2011wz, oliver1, oliver2, 
Daouda:2012ig, Fabris:2012hw}, its application to the primordial universe is still lacking.
\par
It is generally expected that quantum effects may lead to the avoidance of the initial singularity through, for example, a bounce phase which may connect the present expansion phase of the universe to a primordial contracting phase. Such bouncing solutions appear in the context of string theory \cite{copeland, peter, Bronnikov:2002ki} and quantum cosmology \cite{nelson, moniz}.
\par
Loop quantum cosmology is one of the most interesting proposal in the direction of applying quantum gravity to cosmology \cite{rovelli, Ashtekar:2011ni, banerjee, bojo1, date}. One of its main results is the modification of Friedmann's equation leading to,
\begin{equation}
\label{qe}
 H^2 = \frac{8\pi G}{3}\biggr(\rho - \frac{\rho^2}{\rho_c}\biggl)\;,
\end{equation}
where $H$ is the Hubble parameter, $\rho$ is the density, and $\rho_c$ is a constant characteristic density that defines a possible bounce. This last term is absent in the usual Friedmann's equation and formally it corresponds to the effect of negative energy densities. This structure must, in principle, lead to singularity-free model. A class of oscillating, non singular solutions has been determined,
for example, in reference \cite{cai}.
\par
In this paper we intend to study a Rastall's primordial universe model that can mimic the effects represented in Eq.~(\ref{qe}). This can be achieved by considering a specific equation of state in the context of Rastall's theory. We show that bouncing solutions do emerge, but other possible scenarios appear. We will try to classify all possible solutions, by explicit integration of the equations, using a dynamical system approach. This work may be understood as a first step in verifying if Rastall's theory may lead to consistent cosmological scenarios, which can be tested against observations as it has
been done with string effective action \cite{peter}. More fundamentally, such analysis, complemented by a perturbative study, may shed light if indeed Rastall's theory can be considered as a classical, effective formulation of quantum cosmological models resulting from loop quantum cosmology, as we have freely speculated above.
\par
The paper is organized as follows. In Sec.~\ref{Sec:PresProbl}, we present Rastall's cosmology and introduce a suitable equation of state for the primordial phase. In Sec.~\ref{Sec:Intpartcas}, we find explicit solutions for the scale factor and the energy density, for some particular combinations of the parameters of theory which render the equations exactly solvable. In Sec.~\ref{Sec:dynsist}, we perform a dynamical system analysis of the bounce solution. Finally, we present our conclusions in Sec.~\ref{Sec:Concl}.

\section{Basic equations}\label{Sec:PresProbl}

Rastall's theory is based on the following coupled equations \cite{rastall1, rastall2}:
\begin{eqnarray}
\label{re1}
R_{\mu\nu} - \frac{1}{2}g_{\mu\nu}R &=& 8\pi G\biggr(T_{\mu\nu} - \frac{\gamma - 1}{2}g_{\mu\nu}T\biggl)\;,\\
\label{re2}
{T^{\mu\nu}}_{;\mu} &=& \frac{\gamma - 1}{2}T^{;\nu}\;,
\end{eqnarray}
i.e. Einstein equations properly modified in order to account for a non-conservation of the stress-energy tensor (recall that Bianchi identities are still valid). For $\gamma = 1$, GR is recovered.

Under the assumption of homogeneity and isotropy, the space-time is described by the Friedmann-Lema\^{\i}tre-Robertson-Walker metric,
\begin{equation}
ds^2 = dt^2 - a(t)^2\biggr[\frac{dr^2}{1 + kr^2} + r^2(d\theta^2 + \sin^2\theta d\phi^2)\biggl]\;,
\end{equation}
where $a(t)$ is the scale factor function which describes the evolution of the universe, and $k = 0, \pm 1$ is the curvature parameter, indicating a flat, an open or a closed spatial section. Hereafter, we suppose a flat universe $k = 0$ in order to simplify the calculations. Observation also supports a flat or nearly flat universe \cite{komatsu, Hinshaw:2012fq}. Friedmann's and the conservation equation in Rastall's framework read
\begin{eqnarray}
3\frac{\dot{a}^2}{a^2} \ &=& \ 8\pi G \left[\rho - \frac { \gamma - 1}{2}(\rho - 3p) \right]\;, \\
\dot{\rho} + 3\frac{\dot{a}}{a} \left ( \rho + p \right ) \ &=& \  \frac{\gamma - 1}{2} \left ( \dot{\rho} - 3\dot{p} \right )\;,
\end{eqnarray}
where the dot denotes derivation with respect to the cosmic time $t$.
\par
In order to close the above system of equations, it is necessary to define an equation of state. Our main objective is to find bouncing solutions, a behavior at the beginning of the history of the universe which is predicted by some quantum gravity theories, in particular by loop quantum cosmology. Therefore, hereafter we consider the following barotropic equation of state:
\begin{equation}\label{eos}
p \ = \  \alpha\rho + \omega\rho^2\;.
\end{equation}
If $\omega < 0$, this equation of state allows to recover, in the context of Rastall's theory, therefore classically, the structure of the effective Friedmann's equation of loop quantum cosmology. We remark {\it en passant} that such equation of state has a form similar to the so-called modified Chaplygin gas (MCG) models, i.e. $p = A\rho + B/\rho^\alpha$, with $\alpha = -2$. In GR, the MCG models are not very successful \cite{mcg1, mcg2, mcg3, mcg4}, but we show below how richer the possibilities are within Rastall's theory.
\par
With Eq.~\eqref{eos}, Rastall's equations (\ref{re1}, \ref{re2}) assume the following form:
 \begin{eqnarray}
3\frac{\dot{a}^2}{a^2} &=& 8\pi G \left \{ \rho - \frac { \gamma - 1}{2} \left [ (1-3\alpha)\rho - 3\omega\rho^2 \right ] \right \}\;,\\
\dot{\rho} &+& 3\frac{\dot{a}}{a} \left [ (1+\alpha)\rho + \omega\rho^2 \right ] = \frac{\gamma - 1}{2} \left [ (1-3\alpha)\dot{\rho} - 6\omega\rho\dot{\rho} \right ]\;,
\end{eqnarray}
that may be rearranged to give
\begin{eqnarray}
3\frac{\dot{a}^2}{a^2} &=& 4\pi G \left \{ [ (3 - \gamma) + 3\alpha(\gamma - 1) ] \rho + 3\omega(\gamma - 1)\rho^2 \right \}\;, \label{F1}\\
\dot{\rho}  &=&  -6\frac{\dot{a}}{a}\rho\frac{(1 + \alpha + \omega\rho)}{\left [(3 - \gamma) + 3\alpha(\gamma - 1) + 6\omega(\gamma - 1)\rho\right ]} \label{F2}\;.
\end{eqnarray}
If $\alpha \neq -1$, then equation (\ref{F2}) may be integrated in order to give an implicit solution
\begin{equation}
 \rho^{A}(1 + \alpha + \omega\rho)^{\frac{B}{\omega}} = \lambda a^{-6}\;, \label{G1}
\end{equation}
where we have defined
\begin{equation}
A \equiv \frac{(3 - \gamma) + 3\alpha(\gamma - 1)}{1 + \alpha}\;, \qquad \frac{B}{\omega} \equiv \frac{(7\gamma - 9) + 3\alpha(\gamma - 1)}{1 + \alpha}\;, \label{AB}
\end{equation}
and $\lambda$ is a \textit{positive} integration constant.
%In the GR limit ($\gamma = 1$) one has $A = 2/(1 + \alpha) = -\omega/B$, leading to the following evolution of the energy density
%\begin{equation}
%\label{gr1}
%\rho = \frac{(1 + \alpha)\lambda^\frac{1 + \alpha}{2}}{a^{3(1 + \alpha)} - \omega\lambda^\frac{(1 + \alpha)}{2}}\;.
%\end{equation}
If $\alpha = -1$, then Eq.~(\ref{F2}) becomes
% \begin{equation}
% \dot{\rho} \left [ \frac{3 - 2\gamma}{\omega\rho^2} + \frac{3(\gamma - 1)}{\rho} \right ] \ = \  -3\frac{\dot{a}}{a}\;,
% \end{equation}
% and, upon integration, we obtain
\begin{equation}
-\frac{3 - 2\gamma}{\omega\rho} + 3(\gamma - 1)\ln\rho \ = \  -3\ln a + \tilde{\lambda}\;, \label{G2}
\end{equation}
where $\tilde{\lambda}$ is another integration constant.
%When $\gamma = 1$, this expression reduces to $\rho = \omega/(3\ln a - \tilde\lambda)$.
Equations~(\ref{G1}) and (\ref{G2}) can be analytically solved only for some specific cases, namely:
$$
\begin{array}{llll}
(i) & \alpha \neq -1, \; A = \frac{B}{\omega} &\Leftrightarrow& \alpha \neq -1, \; \gamma = \frac{3}{2}\;;\\
(ii) & \alpha \neq -1, \; A = -\frac{B}{\omega} &\Leftrightarrow& \alpha \neq -1, \; \gamma = 1\;;\\
(iii) & \alpha \neq -1, \; A = 0 &\Leftrightarrow& \alpha \neq -1, \; \gamma = \frac{3(\alpha - 1)}{3\alpha - 1}\;;\\
(iv) & \alpha \neq -1, \; \frac{B}{\omega} = 0 &\Leftrightarrow& \alpha \neq -1, \; \gamma = \frac{3(\alpha + 3)}{3\alpha + 7}\;; \\
(v) & \alpha = -1, \; \gamma = \frac{3}{2}\;; & & \\
(vi) & \alpha = -1, \; \gamma = 1\;. & &
\end{array}
$$
In the following section we shall focus on the solution of each of the cases listed above.

\section{Integration of the exactly solvable cases}\label{Sec:Intpartcas}

\subsection{Case $\alpha \neq -1, \; \gamma = 3/2$}\label{Subsec:gamma32}

In this instance we have $A = B/\omega = 3/2$, and Eq.~(\ref{G1}) becomes
\begin{equation}
 \rho(1 + \alpha + \omega\rho) = \lambda a^{-4}\;,
 \end{equation}
whose solution is
\begin{equation}\label{rhocaseA}
 \rho = -\frac{1 + \alpha}{2\omega} \left(1 \pm \sqrt{1 + \frac{4\omega\lambda}{(1 + \alpha)^2} a^{-4}} \right)\;.
\end{equation}
With the above solution, Eq~(\ref{F1}) reduces to
\begin{equation}
(a\dot{a})^2 =  2\pi G \lambda\;. \label{C11}
\end{equation}
Taking the square root and integrating, one obtains
\begin{equation}
a \propto t^{\frac{1}{2}}, \qquad H \propto t^{-1}\;,
\end{equation}
which coincides with expansion behavior of a radiation-dominated universe in standard cosmology. This result is quite curious, because such behavior does not depend on $\alpha$ or $\omega$, which are free parameters.

As for the sign in the parenthesis of Eq.~\eqref{rhocaseA}, considering $\alpha > -1$, there are three possibilities: $\omega > 0$ with the minus sign; $\omega < 0$ with again the minus sign and $\omega < 0$ with the plus sign ($\omega > 0$ with the plus sign would result in a negative energy density). In the first situation, the density decreases with time, tending to zero in the future. In the second situation, the density starts with a finite value and decreases to zero in the future. Finally, in the third situation, the density starts with a finite value and increases with time, tending to a value twice as big as the starting value. Considering $\alpha < -1$, then the density is non-negative only if $\omega > 0$ with the plus sign in the parenthesis of Eq. (16). In such situation, the density decreases with time, tending to a positive value in the future.

\subsection{$\alpha \neq -1, \; \gamma = 1$}

From Eq.~(\ref{AB}) one has that
\begin{equation}
 A = \frac{2}{1 + \alpha}\;, \qquad \frac{B}{\omega} = -\frac{2}{1 + \alpha}\;,
\end{equation}
and Eq.~(\ref{G1}) can be solved obtaining
\begin{equation}
 \rho = \frac{\lambda(1 + \alpha)}{a^{3(1 + \alpha)} - \omega\lambda}\;.
\end{equation}
Since $\gamma = 1$, Eq.~(\ref{F1}) reduces to the standard Friedmann equation:
\begin{equation}
3\frac{\dot{a}^2}{a^2} = \frac{8\pi G\lambda(1 + \alpha)}{a^{3(1 + \alpha)} - \omega\lambda}\;. \label{C21}
\end{equation}
If $\omega < 0$, since the energy density must be positive, we must have $\alpha > -1$ which implies the existence of a singular state $a \to 0$, but with finite energy density $\rho = (1 + \alpha)/|\omega|$. On the other hand, if $\omega > 0$, we have two disconnected evolutions: $i)$ if $\alpha < -1$, then $a^{3(1 + \alpha)} < \omega\lambda$. Therefore, we have an evolution from a singular $a \to 0$ state with finite density $\rho = |1 + \alpha|/\omega$ to a configuration with maximum scale factor $a = (\omega\lambda)^{1/3(1 + \alpha)}$ and diverging density; $ii)$ if $\alpha > -1$, then $a^{3(1 + \alpha)} > \omega\lambda$ and we have an evolution from a minimum scale factor $a = (\omega\lambda)^{1/3(1 + \alpha)}$ and diverging density to an asymptotic Minkowski universe $H \to 0$ for $a \to \infty$.

\subsection{$\alpha \neq -1, \; \gamma = \frac{3(\alpha - 1)}{3\alpha - 1}$}

In this situation, we have that $A = 0$ and, from Eq.~(\ref{AB}), $B/\omega = 6(\gamma - 1)$. Substituting these values in Eq.~(\ref{G1}), we get
\begin{equation}
 \rho = \frac{\lambda}{\omega} a^{-\frac{1}{\gamma - 1}} - \frac{1 + \alpha}{\omega}\;,
\end{equation}
and Friedmann equation (\ref{F1}) becomes
\begin{equation}
 \frac{\dot{a}}{a} = \sqrt{\frac{4\pi G(\gamma - 1)}{\omega}} \, \left[\lambda a^{-\frac{1}{\gamma - 1}} - (1 + \alpha) \right]\;.
\end{equation}
Observe that $\gamma - 1$ and $\omega$ must have the same sign in order to avoid an imaginary Hubble parameter. In particular, $\gamma > 1$ if $\alpha < 1/3$. Remarkably, the above equation can be integrated to give
\begin{equation}\label{scalafcC}
 a = \left [ k\exp \left ( - \sqrt{\frac{4\pi G (1 + \alpha)^2}{\omega(\gamma - 1)}} \, t \right ) + \frac{\lambda}{1 + \alpha} \right ] ^ {\gamma - 1}\;,
\end{equation}
where $k$ is a positive integration constant. It is important to realize that, since $\gamma = 3(\alpha - 1)/(3\alpha - 1)$, here we cannot obtain the GR limit, since it would require $\alpha \to \infty$. There are two main cases:
\begin{enumerate}
 \item If $\alpha < 1/3$, then $\gamma - 1$ and $\omega$ are positive. In this case, the scale factor decreases with time and vice-versa the density increases with time. Note that if $-1 < \alpha < 1/3$, in order to have a positive energy density we must have $a < [\lambda/(1+\alpha)]^{\gamma - 1}$, which is impossible from Eq.~\eqref{scalafcC}. Therefore, the case $-1 < \alpha < 1/3$ is excluded. On the other hand, if $\alpha < - 1$, the scale factor varies from infinity ($t \to -\infty$) to zero at a certain finite time. Correspondingly, the density is always positive and grows from $\rho \to |1 + \alpha|/\omega$ to infinity. This behavior is somehow reverted with respect to the one of standard cosmology.

 \item If $\alpha > 1/3$, then $\gamma - 1$ and $\omega$ are negative. In this case, the scale factor increases with time and vice-versa the density decreases with time. Indeed, the former varies between zero ($t \to -\infty$) to $a = [\lambda/(1+\alpha)]^{\gamma - 1}$ ($t \to +\infty$), where the Hubble parameter vanishes and therefore represents an asymptotic Minkowski universe. Correspondingly, the density varies from $\rho \to (1 + \alpha)/|\omega|$ to zero.
\end{enumerate}

\subsection{$\alpha \neq -1, \; \gamma = \frac{3(\alpha + 3)}{3\alpha + 7}$}\label{Subsec:Bounce}

Here we have $B/\omega = 0$ and, from Eq.~(\ref{AB}), $A = 6(\gamma - 1)$. Thus, Eq.~(\ref{G1}) simplifies to
\begin{equation}
\rho \ = \ \lambda a ^ {- \frac{1}{\gamma - 1}}\;, \label{C41}
\end{equation}
and Friedmann equation (\ref{F1}) can be rewritten as
\begin{equation}
(a^{\frac{1}{\gamma - 1} - 1}\dot{a})^2 \ = \ 4\pi G \lambda (\gamma - 1) \left [ 2(1 + \alpha) a ^ {\frac{1}{\gamma - 1}} + \omega \lambda \right ]\;. \label{C42}
\end{equation}
Hence, the scale factor may be expressed by:
\begin{equation}\label{scalafcD}
a \ = \ \left [ \frac{2\pi G \lambda(1 + \alpha)}{\gamma - 1}t^2 - \frac{\omega\lambda}{2(1 + \alpha)} \right ] ^{\gamma - 1}\;.
\end{equation}
Let us consider some possible scenarios. Again, the GR limit $\gamma \to 1$ is not allowed here.
\begin{enumerate}
 \item If $\alpha < -7/3$, then $\gamma - 1 < 0$. If $\omega > 0$, this corresponds to some sort of ``antibounce'' linking two singularities, since the scale factor varies from zero ($t \to -\infty$) to zero ($t \to +\infty$) with a maximum value $a = \left[\frac{\omega\lambda}{2|1 + \alpha|}\right ]^{\gamma - 1}$ for $t=0$. Correspondingly, the energy density varies from zero ($t \to -\infty$) to zero ($t \to +\infty$), with a maximum value $\rho = 2|1+\alpha|/\omega$ at $t = 0$.

 If $\alpha < -7/3$, but $\omega < 0$, then the range $-\infty < t < +\infty$ is no more allowed because it would correspond to an imaginary scale factor and density. There are two disconnected evolution: In the first $t$ varies from minus infinity to the negative value which nullifies the expression between parenthesis in Eq.~\eqref{scalafcD}, the scale factor varies from zero to infinity and so the energy density. In the second $t$ varies from the positive value which nullifies the expression between parenthesis in Eq.~\eqref{scalafcD} to infinity, the scale factor varies from infinity to zero and so the energy density.
 \item If $-7/3 < \alpha < -1$, then $\gamma - 1 > 0$, but $1 + \alpha$ is negative. If $\omega < 0$ there is no way out, the scale factor and the density become imaginary. Therefore, we can only consider $\omega > 0$. The time $t$ cannot go to $\pm \infty$, but only varies between the roots (finite) of the expression between parenthesis in Eq.~\eqref{scalafcD}. Therefore, the scale factor varies from zero to zero with a maximum $a = \left[\frac{\omega\lambda}{2|1 + \alpha|}\right ]^{\gamma - 1}$ for $t=0$. Correspondingly, the energy density varies from infinity to infinity with a minimum $\rho = 2|1+\alpha|/\omega$ at $t = 0$.
 \item If $\alpha > -1$, then  $\gamma - 1$ and $1 + \alpha$ are positive. If $\omega < 0$, the scale factor varies from infinity ($t \to -\infty$) to infinity ($t \to +\infty$) with a minimum value  $a = \left[\frac{|\omega|\lambda}{2(1 + \alpha)}\right ]^{\gamma - 1}$ and the energy density from zero to zero with a maximum value $\rho = 2(1+\alpha)/|\omega|$. This is a proper bouncing cosmology, connecting two asymptotic Minkowski universes. If $\omega > 0$, we have again two disconnected evolutions. In the first $t$ varies from minus infinity to the negative value which nullifies the expression between parenthesis in Eq.~\eqref{scalafcD}, the scale factor varies from infinity to zero and the energy density from zero to infinity. In the second $t$ varies from the positive value which nullifies the expression between parenthesis in Eq.~\eqref{scalafcD} to infinity, the scale factor varies from zero to infinity and the energy density from infinity to zero.
\end{enumerate}
Note that in order to have a bouncing solution it is necessary a very strict relation between the constants $\alpha$ (which comes from the equation of state) and $\gamma$ (which is a parameter of Rastall's theory). It can be interpreted as a dependence of the violation of the conservation law on the fluid properties. In the scenario of point 3 above, after the bounce the universe starts to expand with a positive second time derivative of the scale factor (i.e. the expansion is accelerated), see Eq.~\eqref{scalafcD}. This is an inflationary phase which resembles power law inflation models. For bouncing models of the universe, inflation is not in general necessary since, for example, there is no causality problem, being the universe eternal. On the other hand, the behavior of quantum fluctuations during this period and their link with the emerging primordial matter power spectrum are complicated issues which deserve more profound investigation. See also \cite{Vitenti:2012cx, Novello:2008ra, Peter:2006hx}.

\subsection{$\alpha = -1, \; \gamma = \frac{3}{2}$}

If $\gamma = 3/2$, then the first term in Eq.~(\ref{G2}) disappears, and thus, it simplifies to
\begin{equation}
\rho \ = \  \lambda a^{-2}\;.
\end{equation}
It is interesting to notice that this equation may be seen as a particular case of Eq.~(\ref{C41}) when $\gamma = \frac{3}{2}$. In spite of that, there is no bouncing behavior in the present situation, since the supposition that $\alpha \neq -1$ was crucial in order to obtain the result in the case $(iv)$. This means that case $(v)$ may be considered as a degenerate form of case $(iv)$.

It should be noticed that, from the results of cases $(i)$ and $(v)$, the scale factor behaves according to $a \propto t^{\frac{1}{2}}$ whenever $\gamma = \frac{3}{2}$, independently on $\alpha$ e $\omega$ (that means, independently on the behavior of the fluid). This fact is even more curious because it coincides with the evolution of the scale factor in standard cosmology for a radiation-dominated universe (this could be explained by the fact that in both cases the trace of the stress-energy tensor is vanishing). However, the density will behave differently, depending on whether $\alpha \neq -1$ or $\alpha = -1$.

\subsection{$\alpha = -1, \; \gamma = 1$}

In this case, Eq.~(\ref{G2}) becomes
\begin{equation}
\rho \ = \ \frac{1}{3\omega\ln a - \omega\tilde{\lambda}}\;,
\end{equation}
and Eq.~(\ref{F1}) reduces to the case of General Relativity,
\begin{equation}
3\frac{\dot{a}^2}{a^2} = 8\pi G\rho = \frac{8\pi G}{3\omega\ln a - \omega\tilde{\lambda}}\;.
\end{equation}
Solving this equation one obtains
\begin{equation}
a = \lambda \exp \left[\left ( \frac{2\pi G}{\omega} \, t^2 \right ) ^{\frac{1}{3}} \right]\;,
\end{equation}
where here $\lambda = e^{\tilde\lambda/3}$.
The expression for the density is
\begin{equation}
\rho = \frac{1}{3\left(2\pi G\omega^2 t^2\right)^{1/3}}\;.
\end{equation}

\section{Dynamical system analysis for the bounce solution}\label{Sec:dynsist}

In the last section we considered some exact solutions of Rastall's cosmology. In the present section we sketch an alternative analysis, via a dynamical system investigation, for the bounce solution. 

Consider the following dynamical system made up of the acceleration equation plus the energy conservation equation in Rastall's cosmology:
\begin{eqnarray}
\label{ds1}
\dot{\rho} &=& -\frac{6H(\rho + p)}{(3 - \gamma) + 3(\gamma - 1)c_{s}^2}\;, \\
\label{ds2}
\dot{H} &=& -4\pi G \left[p + \frac{\gamma - 1}{2}(\rho - 3p) \right] - \frac{3}{2}H^2\;,
\end{eqnarray}
where $c_{s}^2 \equiv \partial p/\partial \rho$ is the speed of sound. Note that Friedmann's equation
\begin{equation}
3H^2 \ = \ 8\pi G \left[\rho - \frac { \gamma - 1}{2}(\rho - 3p) \right]\;, \label{VINC}
\end{equation}
plays the role of a constraint for the dynamical system, indicating which solutions are physical.

\begin{figure}[htbp]
\includegraphics[width=0.5\linewidth]{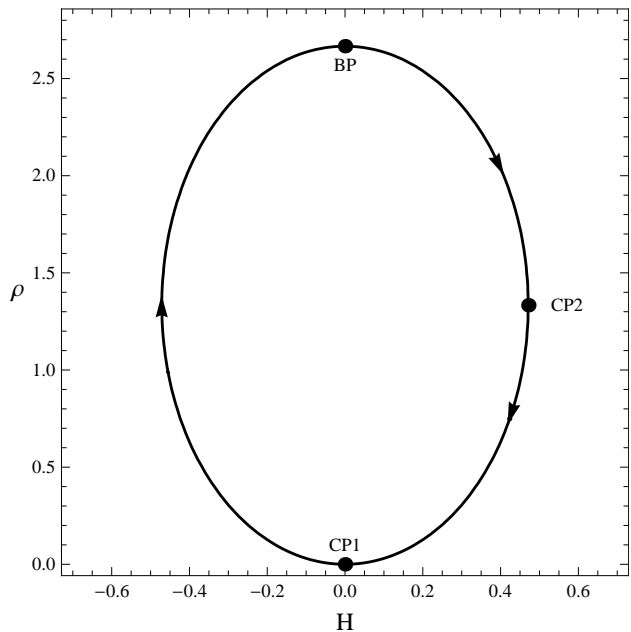}
\caption{Phase space diagram of the bouncing solution of subsection \ref{Subsec:Bounce}, with the choice $\omega = -1$ and $\alpha = 1/3$, the latter implying $\gamma = 5/4$. The points marked are the first critical point CP1 $ = (0,0)$, the second critical point CP2 $ = (\sqrt{2}/3,4/3)$, and the bouncing point BP $ = (0,8/3)$. The trajectory starts from CP1 and ends again in CP1, according to the direction of the arrows.}
\label{Fig1}
\end{figure}

Since in our case we have a fluid of type $p = \alpha \rho + \omega \rho^2$ (which means that $c_{s}^2 = \alpha + 2\omega\rho$), then the dynamical system may be rewritten as
\begin{eqnarray}
\dot{\rho} &=& -\frac{6H\rho(1 + \alpha + \omega\rho)}{(3 - \gamma) + 3(\gamma - 1)(\alpha + 2\omega\rho)}\;, \\
\dot{H} &=& -4\pi G \rho \left[\alpha + \omega\rho + \frac{\gamma - 1}{2}(1 - 3\alpha - 3\omega\rho) \right] - \frac{3}{2}H^2\;.
\end{eqnarray}
The critical points are characterized by the simultaneous conditions $\dot{\rho} = 0$ and $\dot{H} = 0$. In our analysis we shall consider only critical points at finite distance:
\begin{enumerate}
 \item $H = 0\;, \qquad \rho = 0\;,$
 \item $H = 0\;, \qquad \rho = \dfrac{1}{\omega}\dfrac{2\alpha + (\gamma - 1)(1 - 3\alpha)}{3\gamma - 5}\;, $
\item $H = \sqrt{-\dfrac{8\pi G (1+\alpha)(3 - 2\gamma)}{3\omega}}\;, \qquad \rho = -\dfrac{1+\alpha}{\omega}\;.$
\end{enumerate}
The second critical point obeys the constraint (\ref{VINC}) if and only if $\gamma = 3/2$. In this special case, it coincides with the third critical point. Therefore, we consider only the first and third critical points. Remark that, for $\alpha = - 1$, the critical points coincide and represent Minkowski space-time ($H = 0$, $\rho = 0$). Linearizing the dynamical system \eqref{ds1}-\eqref{ds2} one obtains
\begin{eqnarray}
 \delta\dot{\rho} &=& -\frac{6\rho(1 + \alpha + \omega\rho)}{(3 - \gamma) + 3(\gamma - 1)(\alpha + 2\omega\rho)}\delta H\nonumber \\ &+& \left[\frac{36\omega(\gamma - 1)H\rho(1 + \alpha + \omega\rho)}{[(3 - \gamma) + 3(\gamma - 1)(\alpha + 2\omega\rho)]^2} - \frac{6H(1 + \alpha + 2\omega\rho)}{(3 - \gamma) + 3(\gamma - 1)(\alpha + 2\omega\rho)}\right]\delta\rho \;,\\
 \delta\dot{H} &=& -3H\delta H - 4\pi G \left[\alpha + 2\omega \rho + \frac{\gamma - 1}{2}(1 - 3\alpha - 6\omega\rho) \right]\delta\rho\;.
\end{eqnarray}
At the critical point ($\rho = 0$, $H = 0$), the linearized dynamical system reduces to
\begin{eqnarray}
 \delta\dot{\rho} & = &  0\;,\\
\delta\dot{H} & = & - 4\pi G \left[\alpha + \frac{\gamma - 1}{2}(1 - 3\alpha) \right]\delta\rho\;,
\end{eqnarray}
and at $\left[\rho = -(1+\alpha)/\omega, H = \sqrt{-8\pi G(1 + \alpha)(3 - 2\gamma)/(3\omega)}\right]$,
\begin{eqnarray}
\label{lineqdeltarho} \delta\dot{\rho} & = & \frac{6(1 + \alpha)}{(3 - \gamma) - 3(\gamma - 1)(\alpha + 2)} \displaystyle\sqrt{ - \frac{8\pi G(1 + \alpha)(3 - 2\gamma)}{3\omega} }\delta\rho\;, \\
\label{lineqdeltaH} \delta\dot{H} & = & - 3\displaystyle\sqrt{ - \frac{8\pi G(1 + \alpha)(3 - 2\gamma)}{3\omega} }\delta H + 4\pi G \left[\alpha + 2 - \frac{\gamma - 1}{2}(7 + 3\alpha) \right] \delta\rho\;.
\end{eqnarray}
Consider now the case $\gamma = 3(\alpha + 3)/(3\alpha + 7)$, which is the condition (together with $\omega < 0$) for having a bounce, see subsection \ref{Subsec:Bounce}. For the critical point $(\rho = 0, H = 0)$, the linearized system reduces to
\begin{eqnarray}
 \delta\dot{\rho} & = & 0\;,\\
 \delta\dot{H} & = & -4\pi G\frac{(1 + \alpha)(3\alpha + 1)}{3\alpha + 7}\delta\rho\;.
\end{eqnarray}
However, the constraint (\ref{VINC}) implies that in this point $\delta\rho = 0$. So, we can conclude nothing about the behavior of the trajectory through. On the other hand, the denominator of Eq.~\eqref{lineqdeltarho} vanishes, rendering the system singular in the second critical point. In this case, we have to consider another equation. For example, linearizing the constraint (\ref{VINC}), and setting $\gamma = 3(\alpha + 3)/(3\alpha + 7)$, we get:
\begin{equation}
 H\delta H = 8\pi G\frac{1 + \alpha + \omega\rho}{3\alpha + 7}\delta\rho\;,
\end{equation}
which, for $\rho = -(1+\alpha)/\omega$ and $H = (\alpha + 1)\sqrt{-8\pi G/[(3\alpha + 7)\omega]}$, implies that necessarily $\delta H = 0$. Employing this information in Eq.~\eqref{lineqdeltaH}, setting $\gamma = 3(\alpha + 3)/(3\alpha + 7)$, one obtains
\begin{equation}
 \delta\dot{H} = 4\pi G (\alpha + 1)\delta\rho\;.
\end{equation}
This fixes the trajectory in the phase space since, for the second critical point, positive $\delta\rho$ implies that $H$ must grow and vice-versa. That is, the trajectory approaches the critical point from above going through the phase space in a clockwise sense. In \figurename{~\ref{Fig1}} we plot the space diagram of the bouncing solution, via the constraint Eq.~\eqref{VINC}, for the representative choice $\omega = -1$ and $\alpha = 1/3$, the latter implying $\gamma = 5/4$.
% \begin{figure}[htbp]
% \includegraphics[width=0.5\linewidth]{FigNoBounceArrow.eps}
% \caption{Phase space diagram of the solution with no bounce, obtained for $\gamma = 3/2$, $\omega = -1$ and $\alpha = 1/3$. The points marked are the first critical point CP1 $ = (0,0)$, the second critical point CP2 $ = (4/3,0)$, both for $H = 0$, which implies the absence of a bounce.}
% \label{Fig2}
% \end{figure}
% In \figurename{~\ref{Fig2}} we plot, as another example, the phase space for the solution found in subsection \ref{Subsec:gamma32}, for $\gamma = 3/2$. In this case, the linearized system, using again the constraint, becomes
% \begin{eqnarray}
%  \delta\dot{\rho} & = &  0\;,\\
% \delta\dot{H} & = & 0\;,
% \end{eqnarray}
% for both the critical points. In this case we cannot conclude anything about the nature of the two points. However, knowing the solution of \ref{Subsec:gamma32}, we can place the initial point of the trajectory in any place on the curve of \figurename{~\ref{Fig2}} and follow the evolution according to the arrows.

\section{Conclusions}\label{Sec:Concl}

In this paper, we study a classical cosmological model based on Rastall's theory of gravity that can mimic quantum effects typical of loop quantum cosmology. Through the introduction of a convenient equation of state, see Eq.~\eqref{eos}, we reproduce repulsive effects as the singularity is approached during the evolution. We also obtain some exact solutions, one of which exhibits a bouncing behavior, described in the subsection \ref{Subsec:Bounce}. In order to obtain such bouncing scenario, a connection between the equation of state parameter $\alpha$ and Rastall's parameter $\gamma$ must take place. Besides, some interesting radiation-like (even without radiation!) scenarios free of curvature singularities emerge. We finally perform a dynamical system analysis confirming these results.
\par
The bouncing solution here obtained can be studied perturbatively using the full covariant equations. This may allow to establish some observational constraints on the model. More important, such perturbative analysis may help to verify if the similarity between Rastall's theory and loop quantum cosmology is just an accident or may contain a deeper meaning. This is an important aspect of the problem, since the basis of Rastall's theory contains some ingredients of quantum effects in curved space-time.
\par
It must be stressed the some perturbative studies in the context of loop quantum cosmology have been already carried out \cite{bojo2, bojo3, lqc1, lqc2, lqc3}. But, most of them focused an inflationary behavior and in \cite{lqc3} the authors consider gravitational waves in a bouncing scenario. It will be crucial to confront the results of these references with a full perturbative analysis of the Rastall's models exhibited here. We hope to address this question in the future.

\acknowledgements

We thank FAPES (Brazil) and CNPq (Brazil) for partial financial support. We thank the anonymous referee for his remarks on our text.

\end{document}